# Alpha Wavelet Power as a Biomarker of Antidepressant Treatment Response in Bipolar Depression

Wojciech Jernajczyk[1], Paweł Gosek[2,3], Miroslaw Latka[4], Klaudia Kozlowska[4],

Łukasz Święcicki[2], Bruce J. West[5]

[1] Department of Clinical Neurophysiology, Institute of Psychiatry and Neurology, Warsaw, Poland.

[2] 2nd Psychiatric Clinic, Department of Affective Disorders, Institute of Psychiatry and Neurology, Warsaw, Poland.

[3] Department of Forensic Psychiatry, Institute of Psychiatry and Neurology, Warsaw, Poland.

[4] Department of Biomedical Engineering, Wroclaw University of Science and Technology, Poland.

[5] Mathematics and Information Science Directorate, Army Research Office, Durham, NC, USA.

# Abstract

There is mounting evidence of a link between the properties of electroencephalograms (EEGs) of depressive patients and the outcome of pharmacotherapy. The goal of this study was to develop an EEG biomarker of antidepressant treatment response which would require only a single EEG measurement. We recorded resting, 21-channel EEG in 17 inpatients suffering from bipolar depression in eyes closed and eyes open conditions. The EEG measurement was performed at the end of the short washout period which followed previously unsuccessful pharmacotherapy. We calculated the normalized wavelet power of alpha rhythm using two referential montages and an average reference montage. The difference in the normalized alpha wavelet power between 10 responders and 7 nonresponders was most strongly pronounced in link mastoid montage in closed–eyes condition. In particular, in the occipital (O1, O2, Oz) channels the wavelet power of responders was up to 84% higher than that of nonresponders. Using a novel classification algorithm we were able to correctly predict the outcome of treatment with 90% sensitivity and 100% specificity. The proposed biomarker requires only a single EEG measurement and consequently is intrinsically different from biomarkers which exploit the changes in prefrontal EEG induced by pharmacotherapy over a given time.

# 1. Introduction

Bipolar depression (BD) remains a major unresolved challenge for psychiatric therapeutics. Several longitudinal studies indicate that the proportion of total time in the depressive component of BD is far greater than in manic or hypomanic phases (Frye et al. 2014). The recent study (Tundo et al. 2014) and the meta-analysis (Vázquez et al. 2013) demonstrated that the overall efficacy of pharmacotherapy for BD was significantly greater with antidepressants than with placebo-treatment and not less than that of unipolar depression (UD). Moreover, the risk of non-spontaneous mood-switching induced by antidepressant treatment seems to have been overestimated.

However, low remission rate and long response time are two fundamental difficulties associated with antidepressant treatment. These difficulties were documented in the STAR*D study in which 47% of the 2876 major depressive disorder (MDD) outpatients responded to citalopram but only 28% achieved full remission (Trivedi et al. 2006). Taking into account that clinical improvement may occur as long as 8 weeks after the onset of treatment (Bauer et al. 2007), it is not surprising that the selection of an antidepressant is often based on lengthy trial-and-error. This difficulty has driven the search for treatment response biomarkers for several decades.

Biomarkers rooted in electroencephalography (EEG) are particularly appealing, since this technique is both cost effective and readily available in clinical practice (Baskaran et al. 2012; Iosifescu 2011). The efficacy of such biomarkers was almost exclusively investigated in pharmacotherapy of major depressive disorder. The very first qualitative analysis of EEG in MDD patients (Ulrich et al. 1986; Ulrich et al. 1984) suggested the possibility that the Fourier spectrum of pre-treatment EEG may in fact reflect patients' propensity for pharmacological treatment. During the pharmacotherapy, the alpha power decreased in responders and increased in nonresponders. However, the observed differences were not statistically significant. In the later studies of (Knott et al. 2000; Knott et al. 1996) and (Bruder et al. 2001) there was no significant difference in the baseline alpha power between responders and nonresponders. Interestingly enough, (Bruder et al. 2001) found a gender-specific effect: the alpha power asymmetry in the *open* eyes condition strongly differentiated female responders from female nonresponders. In more recent work, (Bruder et al. 2008), the comparison of the natural logarithm of alpha power in the occipital region (O1,O2) in the mixed (open and closed eyes) condition yielded $p=0.06$ slightly greater than the conventional threshold of significance. However, the responders' log power asymmetry was greater than that of nonresponders ($p=0.02$). (Bruder et al. 2008) rekindled interest in alpha waves measures in MDD by predicting treatment response using the log alpha power (sensitivity 72.7%, specificity 57.5%) and alpha asymmetry (sensitivity 63.6%, specificity 71.4%).

In the eyes-closed condition, the prominent alpha wave spindles are the hallmark of the *non-stationarity* of EEG signals (Shaw 2003). Nevertheless, the Fourier analysis, whose applicability requires the signal to be stationary, is predominantly used in quantitative EEG. In healthy individuals, the alpha power spectrum is usually stable but like other EEG traits must not be considered as unchangeable. However, intersubject differences are high, which was demonstrated as early as in 1934 by Adrian and Matthews (Niedermeyer 2005). As long as the prediction of treatment response is based upon the changes in a patient's spectrum that are manifested shortly after the initiation of pharmacotherapy (Leuchter et al. 2009b; Leuchter et al. 2009c), the intersubject variability is irrelevant. However, the effectiveness of any prediction algorithm based on a *single* EEG measurement may be degraded by the intersubject variability of alpha rhythm. In this work we test the hypothesis that in bipolar depression the differences in alpha power topography between responders and nonresponders may be used to develop a clinically efficient biomarker of antidepressant treatment response once nonstationarity and intersubject variability are taken into account.

Routine inpatient psychiatric care encompasses both patients who did not take an antidepressant for an extended period of time and those who were admitted because of a failure of ongoing pharmacotherapy. In the latter group, in sharp contrast to research trials, the length of the washout which precedes the change of antidepressant should be as short as possible. In the patient's interest, the washout period is determined only by the properties of the drugs involved in the treatment. Herein, we present a way to predict the outcome of pharmacotherapy in a diverse cohort of BD patients using EEG measured at the end of a washout period which is so short that the influence of unsuccessful pharmacotherapy on alpha rhythm may not be excluded a priori. We hypothesize that the prediction is also effective in this case.

## 2. Materials and Methods

### 2.1 Patients

The study protocol was approved by the Bioethical Commission of the Institute of Psychiatry and Neurology. The study comprised 26 depressive bipolar inpatients from the Institute of Psychiatry and Neurology in Warsaw. All of them were right-handed, aged between 18 and 75 years, and met International Classification of Diseases ICD-10 criteria of bipolar depression. All patients received a written description of the protocol and signed the informed consent form. Subjects were excluded if they met the criteria for substance abuse, were pregnant, had psychotic depression, organic brain pathology (confirmed by MRI or CT scan), a history of chronic benzodiazepine use, suffered from severe neurological disorders (e.g. epilepsy, Alzheimer's or Parkinson's disease) or general medical conditions. We enrolled only patients with a history of unchanged normothymic treatment during 4 weeks before the trial. Drug selection was based on initial psychiatric status, previous treatment history and patient preference. Antidepressive monotherapy was preceded by short washouts of an average length of $55 \pm 31$ h. The study was scheduled for 4 weeks of active treatment.

Out of 26 subjects who entered the study, 17 reached completion and 9 patients left the trial due to: worsening of the depressive symptoms (n=2), hypomania (n=2) medication intolerance (n=1) or other reasons (n=1). Three subjects requested a discharge from the hospital before the end of the trial. Out of 17 subjects 10 responded to treatment. There was no difference in pre-treatment severity of depression on the Montgomery-Åsberg Depression Rating Scale (MADRS) and Beck Depression Inventory (BDI) between responders and nonresponderss. The clinical characteristics of both groups are presented in Table 1.

The length of the wash-out period did not differ significantly in both groups ($53 \pm 31$h for responders and $61 \pm 37$h for nonresponderss). For the patients who completed the study the antidepressant selection was as follows: venlafaxine (n=6), bupropion (n=4), citalopram (n=3), reboxetine (n=2), fluoxetine (n=1) and mirtazapine (n=1). Doses of the antidepressants were consistent with official product characteristics (SPC). In the responders group (n=10) patients received treatment with: bupropion (n=3), venlafaxine (n=3), reboxetine (n=2), citalopram (n=1), fluoxetine (n=1) and in the nonresponderss group with: venlafaxine (n=3), citalopram (n=2), bupropion (n=1) and mirtazapine (n=1).

Normothymic treatment was unchanged during the trial. In the responders group 5 subjects received monotherapy: lamotrigine (n=3), olanzapine (n=1), lithium (n=1) and 5 received a combination of normothymics: lithium + lamotrigine (n=2), lithium + quetiapine (n=1), lithium + lamotrigine + quetiapine (n=1), lamotrigine + carbamazepine + olanzapine (n=1). In the nonresponderss group two subjects were treated with monotherapy: one with olanzapine and one with quetiapine. Five subjects received a combination of normothymics: lithium + olanzapine (n=1), lithium + lamotrigine (n=1), lithium + valproate (n=1), lithium + carbamazepine + olanzapine (n=1), lamotrigine + olanzapine (n=1).

## 2.2 Assessment of Depressive Symptoms

Depressive symptoms were quantified by the MADRS, administered by the attending physician, and the BDI which was completed by patients. The assessment of depressive symptoms was done at baseline and day 28 of the trial. Response to treatment was defined as the reduction of the final MADRS score by more than 50%. A final MADRS score less than or equal to 10 corresponded to remission.

## 2.3 EEG Recording

The EEG recording was done at baseline, day 7 and 28 of the trial. In this work we analysed the baseline recording. The 10-20 international standard was used to position 21 Ag/AgCl electrodes (impedances were below 5 kΩ). The ground electrode was placed between Fpz and Fz. Two referential montages were used: the conventional linked mastoid (LM) and a REF montage for which the reference electrode was mounted between Fz and Cz. The EEG was recorded through a Grass Telefactor Comet data acquisition system with the sampling frequency of 200 Hz and bandpass of 0.3–70 Hz. The EEG waveforms corresponding to two referential montages were recorded simultaneously. We also generated the average reference (ARE) montage using as a common reference the instantaneous average of all 21 electrodes. Subjects remained in supine position in a quiet room. The measurement consisted of 3 five-minute intervals. During the first interval, subjects had open eyes. The eyes-closed intervals were separated by a short (approximately 10 s) blinking interval.

## 2.4 Data Analyses

The continuous wavelet transform of signal $s(t)$, such as EEG record, is defined as:

$$W_s(a, t_0) = \frac{1}{\sqrt{a}} s(t) \psi^* \left( \frac{t-t_0}{a} \right) dt \quad (1)$$

(Latka et al. 2005; Latka et al. 2003) and references therein. In the above formula $a$ is the scale and $t_0$ indicates the localization of the wavelet. We refer to the square of the complex modulus of $W_s$ as the wavelet power. In this work we will use the wavelet power averaged over time interval:

$$w(f) = \langle |W_s(f, t_0)^2| \rangle_{t_0}. \quad (2)$$

The dual localization of wavelets in time and frequency enables us to associate a pseudo-frequency $f_a$ with the scale $a$

$$f_a = \frac{f_c}{a\delta t} \quad (3)$$

where $f_c$ is the center frequency and $\delta t$ is the sampling period of the signal $s(t)$. Thus, the value of wavelet coefficient reflects the *local* properties of the signal at given scale (pseudofrequency). From the plethora of existing mother functions one should judiciously choose one that is effective in extracting these features of the signal that are important for the problem at hand. Herein, we employ complex Morlet

$$\Psi(t) = \frac{1}{\sqrt{\pi f_b}} e^{i2\pi f_c t} e^{-t^2/f_b} \quad (4)$$

where center frequency $f_c$ and bandwidth parameter $f_b$ may be independently adjusted. In Figure 1 we present the time averaged wavelet power $w(f)$ of a monochromatic wave with frequency 10 Hz (value close to the average frequency of alpha waves in healthy adult subjects) plotted as a function of wavelet transform pseudofrequency $f_a$. It is apparent that for $f_c=1.8$ and $f_b=1$ the width of the wavelet power distribution essentially covers the alpha band (8-13 Hz). For this choice of the complex Morlet parameters and pseudofrequency $f_a=10$ Hz the wavelet power is just the weighted average of power in the entire alpha band. In other words, wavelet smoothes out the alpha band spectrum. The use of a just single pseudofrequency to characterize the power in the alpha range is not by all means obvious. In this work we are interested in the topography of the alpha wave power. Therefore, we normalize power $w(f;i)$ in the *i-th* EEG channel by the total power

$$n(f;i) = \frac{w(f;i)}{\sum_{i=1}^{21} w(f;i)}. \quad (5)$$

Even if the dominant alpha frequency of the subject is different that the chosen value of 10 Hz, the wavelet power is approximately proportionally reduced in all channels, preserving the topography of the normalized wavelet power. Please note that the dominant alpha frequency, averaged over all channels and patients, was equal to 9.7 Hz for responders and 9.3 Hz for nonresponders. Moreover, there were no statistically significant differences between responders and nonresponders in any of the analyzed channels.

Many groups of authors advocate the use of a narrow Fourier band centered around the dominant alpha frequency to characterize resting state or task-related changes in alpha rhythm (Klimesch 1999; Klimesch 1997). In Fig. 1 we provide an example of Fourier spectrum of a depressive patient with broad distribution of alpha power without a distinctive dominant frequency. These traits of Fourier spectrum are common in patients and motivated us to use a broad analysing wavelet.

Log transformation is frequently used in analysis of physiological data. The question arises as to whether logarithm of wavelet power should be used in Equation (5). The justification of such transformation is the assumption that susceptibility to antidepressant treatment is multiplicatively related to EEG alpha wavelet power. Although there is no a priori justification of such relation, the application of log transformation is a viable modification of the presented prediction algorithm.

In the eyes-open condition, we calculated alpha wavelet power for contiguous EEG data segments without manual or software excising of eye blinks. Therefore, it is worth mentioning that the average number of blinks per minute in the eyes-open interval was similar for responders and nonresponders ($28 \pm 15$ vs. $25 \pm 21$).

A neurophysiologist selected a 2 min data segment from the eyes-open interval which, apart from eye blinks, was free from artifacts. The 2 min artifact-free EEG segment was also extracted from the first eyes-closed interval. Until the end of study, neither the neurophysiologist nor the persons who performed data analyses had access to patients' treatment records. For both referential montages (LM and REF) and average reference we calculated the continuous wavelet transform using the Morlet mother function with parameters $f_c=1$ and $f_b=1.8$. The calculations were performed for the pseudofrequency 10 Hz. The wavelet power over the entire data segment was averaged for each EEG channel. Finally, the averaged wavelet power in each channel was normalized by the total averaged wavelet power from all 21 channels. Consequently, the normalized wavelet power is independent of the subject's EEG amplitude. Wavelet transforms, in stark contrast to traditional Fourier methods, are intrinsically more robust with respect to eye movement or blink artifacts. This property allowed us to calculate alpha wavelet power for contiguous EEG data segments without manual or software excising of eye blinks.

The frequency of a peak value of wavelet power in the interval 8-13 Hz is referred to as a dominant alpha wave frequency.

We define an alpha power ratio as the ratio of the sum of alpha wavelet power at frontal (Fp1, Fp2, Fpz) sites to the sum of alpha power at occipital (O1, O2, Oz) sites.

The Mann–Whitney U test was used to assess the statistical significance of differences in normalized wavelet power $n(10\ Hz)$ and response index between responders and nonresponders. In all cases the traditional $p=0.05$ was chosen as the threshold of statistical significance.

## 2.5 Prediction of Antidepressant Treatment Response

From the mathematical point of view, prediction of treatment response is equivalent to binary classification based upon a single criterion (such as normalized alpha wavelet power or alpha power ratio). The receiver operating characteristic (ROC) provides a rigorous framework for such classification (Hanley 1989). This framework enables the determination of the optimal classification threshold and a qualitative assessment of statistical significance. The area under the receiver operating characteristic curve (AUROC) was used to quantify the performance of the binary classifier. When classification was feasible we calculated the optimal threshold value (cut-off point) as well as the sensitivity, specificity, and accuracy.

It turns out that regardless of the chosen montage prediction of antidepressant treatment, response is usually possible at several EGG sites. Therefore, we elected to test the prediction algorithm based on a response index – the percentage of channels in which a patient was classified as a responder. In other words, a patient is classified as a responder when this index is greater than 50%; otherwise, a patient is assigned to the nonresponder category. At first, such a choice seems arbitrary. However, the proposed classification scheme is reminiscent of the nearest neighbor pattern classification introduced by (Cover and Hart 1967). The existence of two classes (responders and nonresponders) leads to the classification threshold equal to 50%. Ideally, such an index should take on the value of 100% for responders and 0% for nonresponders.

# 3. Results

## 3.1 Average Reference Electrode (ARE) Montage

In Fig. 2 we present the topography of alpha wavelet power for nonresponders (A and D) and responders (B and E) in the eyes-open (EO) and eyes-closed (EC) conditions. Figs. 2c and 2f show the relative difference of the wavelet power between responders and nonresponders (relative to nonresponders) for open and closed eyes, respectively. In these two figures, the red thick circles around EEG site labels indicate channels for which the AUROC was significantly greater than 0.5.

For open eyes, the AUROC was significantly greater than 0.5 in five channels listed in Table 2. The largest value, 0.84, occurred at the C3 site. The wavelet power of responders $nR(10Hz; C3) = 0.021 \pm 0.005$ was smaller than that of nonresponders $nN(10Hz;C3) = 0.029 \pm 0.007$ and this difference was statistically significant ($p=0.02$). For the cut-off threshold 0.024 the prediction of antidepressant treatment response had 82% accuracy, 80% sensitivity and 86% specificity.

For closed eyes, the classification was feasible in 8 channels (Table 2). The AUROC took on the highest value, 0.87, in the Fpz channel. The classification for this channel had 82% accuracy, 80% sensitivity and 86% specificity. The wavelet power of responders $nR(10Hz; Fpz) = 0.037 \pm 0.006$ was smaller than that of nonresponders $nN(10Hz;Fpz) = 0.05 \pm 0.02$ ($p=0.001$).

The response index of responders was notably higher than that of nonresponders both for open (76% ± 26% vs. 20% ± 20%, $p=0.0005$) and closed (76% ± 22% vs. 23% ± 19%, $p=0.001$) eyes (Table 5). The value of the index averaged over both conditions was equal to 76% ± 20% and 21% ± 17% for responders and nonresponders, respectively ($p=0.0004$).

For closed eyes, the alpha power ratio for responders $0.30 \pm 0.06$ was smaller than that of nonresponders $0.41 \pm 0.13$ ($p=0.03$). For open eyes, the difference was not statistically significant.

## 3.2 Link Mastoids (LM) Montage

For open eyes, the AUROC was significantly greater than 0.5 in five channels listed in Table 3. The largest value, 0.79, occurred at Fz site. The wavelet power of responders $nR(10Hz; Fz) = 0.059 \pm 0.008$ was smaller than that of nonresponders $nN(10Hz;Fz) = 0.064 \pm 0.007$ and this difference was not statistically significant ($p=0.06$). For the cut-off threshold 0.056 the prediction of antidepressant treatment response had accuracy of 82%, sensitivity 70% and specificity 100%.

For closed eyes, the classification was feasible in 9 channels (Table 3). The AUROC took on the highest value, 0.86, in O1 channel. The wavelet power of responders $nR(10Hz; O1) = 0.07 \pm 0.03$ was greater than that of nonresponders $nN(10Hz; O1) = 0.039 \pm 0.013$ ($p=0.001$). The classification for this channel had accuracy of 82%, sensitivity 80% and specificity 86%.

The response index of responders was notably higher than that of nonresponders both for eyes-open (80% ± 23% vs. 20% ± 28%, $p=0.002$) and eyes-closed (76% ± 22% vs. 23% ± 20%, $p=0.001$) conditions (Table 5). The value of the index averaged over both conditions was equal to 76% ± 20% and 21% ± 17% for responders and nonresponders, respectively ($p=0.0004$).
The values of indices averaged over both conditions were exactly the same as those for the ARE montage.

For closed eyes, the alpha power ratio for responders 0.85 ± 0.42 was smaller than that of nonresponders 1.81 ± 1.3 (*p*=0.04). As in the case of the ARE montage, for open eyes, the difference was not statistically significant.

### 3.3 REF Montage

For this montage, the AUROC was significantly greater than 0.5 in only three channels in eyes open condition: F4, C4 and C3 (Table 4). The highest value of AUROC, 0.76, was observed in channel F4. The wavelet power of responders $nR$(10Hz; F4) =0.03 ± 0.01 was smaller than that of nonresponders $nN$(10Hz; F4) =0.04 ± 0.01. However, this difference was not statistically significant (*p*=0.08). The classification for this channel had 88% accuracy, 100% sensitivity and 71% specificity. The response index of responders was higher than that of nonresponders: 83% ± 18% vs. 23% ± 25% (*p*=0.0006).

In both the open and closed eyes condition there was no difference in alpha power ratio between responders and nonresponders.

### 3.4 Prediction of Treatment Response

The values of the response index for all patients are collected in Table 5. The index was calculated for open and closed eyes for both the ARE and LM montages. Fig. 3 visualizes the application of these indices to prediction of antidepressant treatment response (white and black boxes represent assignment to responders (R) and nonresponders (N), respectively). For both ARE and LM montages, the prediction based on the response index averaged over the eyes-open and eyes-closed conditions resulted in only one misclassification (90% sensitivity, 100% selectivity, and 94% accuracy). However, it was not the same subject that was misclassified. The prediction based on the eye-open response index for the REF montage resulted also in one misclassification (100% sensitivity, 86% selectivity, and 94% accuracy).

## 4. Discussion

With the exception of the pilot study of (Bares et al. 2012), the efficacy of EEG biomarkers of antidepressant response was exclusively investigated in the pharmacotherapy of MDD. Unipolar depression is considered to be a disorder of right hemispheric functions (particularly these associated with the temporoparietal region) or of interaction between the hemispheres with relative right-sided or nondominant impairment (Small 2005). In the broader perspective, the dominant and nondominant hemispheres subserve positive and negative affects, respectively (Debener et al. 2000) and references therein. Therefore, EEG asymmetry indices seem to be a natural choice for the prediction of the outcome of antidepressant treatment (Ulrich et al. 1984). A study of 50 MDD patients treated with fluoxentine (Bruder et al. 2001) demonstrated that in the *eyes-open* condition the difference in overall alpha asymmetry (averaging was done over homologous sites of the anterior, central, and posterior regions) between responders and nonresponders was significant only for females, not for males. In later work, (Bruder et al. 2008) using the *occipital* alpha asymmetry of resting EEG (mixed open and closed eyes scenario) predicted the outcome of 12 week fluoxentine treatment with 63.6% sensitivity and 71.4% specificity. (Tenke et al. 2011; Bruder et al. 2008) established that in MDD responders had greater alpha power compared with nonresponders and healthy control subjects, with the largest differences at occipital sites O1 and O2 where alpha rhythm was most strongly pronounced. (Bruder et al. 2008) using alpha rhythm power achieved 72.7% sensitivity and 57.5% specificity in prediction of antidepressant treatment response. By combing two metrics: asymmetry and power of alpha waves, they improved the prediction performance to 83.3% sensitivity and 67.7% specificity. (Tenke et al. 2011) reported 92% specificity, albeit accompanied by 50% sensitivity.

The difference in the normalized alpha wavelet power between responders and nonresponders was most strongly pronounced in LM montage in the closed–eyes condition. In particular, in the occipital (O1, O2, Oz) channels the wavelet power of responders was up to 84% higher than that of nonresponders (*cf.* Table 3), in agreement with the results of the previous studies (Tenke et al. 2011; Bruder et al. 2008). Moreover, the ratio of the wavelet power in the frontal channels (Fp1, Fp2, Fz) to that of the occipital channels was significantly higher ($p=0.04$) in nonresponders ($1.8 \pm 1.2$) than in responders ($0.9 \pm 0.4$). The same traits of alpha wavelet power topography were observed in ARE montage in the closed-eyes condition. It is worth pointing out that for both LM and ARE montages for closed eyes the wavelet power of responders in frontal, central, and parietal electrodes was smaller in responders.

The performance of a single-channel prediction based on the *normalized* alpha wavelet power matched or exceeded those of previous studies. For example, for the ARE montage (Table 2) in the eyes-open condition, out of five channels for which discrimination between responders and nonresponders was possible (AUROC statistically greater than 0.5), for three the classification accuracy was equal to 82%. For eyes closed, accuracy was equal to 82% for three out of eight channels. The performance of classification for the LM montage (Table 3) was comparable. For the REF montage (Table 4), prediction of treatment response was possible only in the eyes-open condition at three EEG sites. The classification based on alpha power ratio was possible for closed eyes for both the ARE (82% accuracy, 80% sensitivity, 85% specificity) and LM (76% accuracy, 70% sensitivity, 85% specificity).

We elected to test the prediction algorithm in which the classification of a patient is based on the response index (percentage of channels for which patient is classified as responder). We can see in Fig. 3 that the prediction for open eyes is more accurate than that for closed eyes, both for the ARE (94% vs. 82%) and LM (82% vs. 76%) montages (Table 5). Averaging of the response index over both conditions led to only one misclassification (94% accuracy, 90% sensitivity 90%, and 100% selectivity) for both montages. While the ARE montage seems to be the most suitable for the prediction of treatment response the prevalent link mastoids also leads to acceptable classification performance. It should be emphasized that the influence of the choice of reference and montage on the outcome of antidepressant treatment prediction in unipolar depression has not yet been thoroughly investigated. We believe that the presented results will facilitate development of effective EEG biomarkers of antidepressant response in bipolar depression.

Wavelet alpha power may be used to define three potential biomarkers of antidepressant treatment response: normalized power at a given site, response index, and ratio of frontal to occipital power. The evaluation of their efficacy and the selection of the best biomarker requires further clinical studies.

To our knowledge this is only the second study which explores the possibility of predicting response to antidepressant intervention in bipolar affective disorder. In the recent work (Bares et al. 2012) found that the treatment response in BD patients is associated with the reduction of prefrontal *theta* cordance after one week following administration of a new antidepressant. Such a reduction was first observed in UD (Bares et al. 2010; Leuchter et al. 2009a; Cook et al. 2009; Bares et al. 2008; Bares et al. 2007; Cook et al. 2005; Cook et al. 2002; Cook et al. 1999).

As we pointed out in the introduction, previous attempts to employ alpha waves for predicting the outcome of pharmacotherapy of unipolar depression have met with limited success. Nevertheless, the properties of this EEG band have been incorporated into the Antidepressant Treatment Response Index (ATR) that combines *prefrontal* EEG theta and alpha power from baseline and after a week of pharmacotherapy (Leuchter et al. 2009b; Leuchter et al. 2009c). However, the algorithm behind ATR is proprietary and cannot be independently verified. In their recent study (Leuchter et al. 2009c) forecast response to escitalopram with 58% sensitivity and 91% specificity.

Nearly half of the depressive patients do not respond to initial antidepressant treatment. (Cipriani et al. 2009) performed multiple-treatment meta-analysis of 117 randomized controlled trials (25928 participants) to assess the efficacy of 12 new-generation antidepressants. Mirtazapine, escitalopram, venlafaxine, and sertraline turned out to be the most efficacious. With the exception of reboxetine, the reported differences were rather moderate (odds ratios in binary comparisons were of the order of 1.3). The usual strategy is either to switch medications or add a drug with a different mechanism of action (MOA) (Stahl and Grady 2003). However, it has never been proven that MOA enhances the effectiveness of switching or combining antidepressants (Thase and Rush 1997). On the contrary, the results of level II treatment in STAR*D indicate that response or remission is independent of MOA (Rush et al. 2006). Herein we demonstrated that a highly effective prediction of response to antidepressants in BD is not affected by ongoing pharmacotherapy, which is corroborated by previous studies with and without a short wash-out period (Cook et al. 2005). In other words, the prediction can be based on a *single* EEG measurement which can be taken either prior to the onset of pharmacotherapy or during the washout period which may precede the change of medication. It is worth emphasizing that the presented approach is intrinsically different from the prediction based on the changes in prefrontal EEG induced by pharmacotherapy over a given time (usually a week) and quantified either by ATR index (Iosifescu et al. 2009; Leuchter et al. 2009b; Leuchter et al. 2009c) or theta band cordance.

The nature of differences of alpha waves observed between responders and nonresponders is not fully understood. (Bruder et al. 2008; Bruder et al. 2005) argue that in unipolar depression these properties reflect endophenotypic vulnerability to depression, while others support the hypothesis of time-dependent susceptibility of depressive patients to pharmacotherapy. The latter hypothesis has been expressed in the literature in a variety of implicit forms for quite some time. Either of the two hypotheses can be ultimately verified only with a QEEG metric that proves highly successful in the prediction of antidepressant treatment response. We strongly believe that any such metric should take into account at least two fundamental features of human EEG time series: non-stationarity and intersubject variability. Herein we pointed to the mathematical framework of continuous wavelet transform as a possible source of such metrics. The limitations of this and similar studies (Iosifescu 2011) are related to open, nonrandomized treatment with a variety of medications. There is no doubt that the rigorous testing of the presented approach to prediction of antidepressant treatment response on a much larger cohort of depressive patients is required before a definitive assessment of its applicability can be made.

Relatively little is known about the differences between the properties of brain oscillations in unipolar and bipolar depression. (Tas et al. 2014; Lee et al. 2010; Lieber 1988). Thus, the question arises as to whether the presented algorithm is applicable to unipolar depression and whether it can be modified to predict antidepressant response in a cohort of unipolar and bipolar patients. The latter question is particularly significant, since previous studies demonstrated that 60% of BD cases were incorrectly diagnosed as UD and consequently were inappropriately treated (Goodwin and Jamison 2007; Dunner 2003). These questions are the subject of our ongoing research.

# Acknowledgments

This research was supported by an Intramural Grant from the Institute of Psychiatry and Neurology in Warsaw. The authors did not have any financial disclosures or potential conflicts of interest.

**Corresponding author:** Miroslaw Latka, Institute of Biomedical Engineering, Wroclaw University of Science and Technology, Wybrzeże Wyspiańskiego 27, 50-370 Wrocław, Poland.

**Email:** Miroslaw.Latka@pwr.edu.pl

**Phone:** +48606635331

**Fax:** +48713277727

Table 1 Clinical characteristics of responders and nonresponders.

|  | Responders (n=10) | Nonresponderss (n=7) |
|---|---|---|
| Gender | F=8; M=2 | F=6; M=1 |
| Age (years) | 44.4 ± 20.1 | 50.0 ± 13.3 |
| Wash-out period (h) | 53.2 ± 31.2 | 60.7 ± 36.7 |
| MADRS pre-treatment | 28.6 ± 6.4 | 29.6 ± 8.5 |
| MADRS post-treatment | 6.4 ± 5.4 | 22.4 ± 6.9 |
| BDI pre-treatment | 37.6 ± 11.7 | 34.1 ± 11.7 |
| BDI post-treatment | 7.0 ± 5.6 | 25.7 ± 11.7 |

Table 2 The outcome of binary classification (prediction of treatment response) based on the normalized alpha wavelet power for eyes-open (EO) and eyes-closed (EC) condition. The power was calculated for the ARE montage. Only the channels for which the AUROC was significantly greater than 0.5 are presented. Δ is the relative percentage difference of the wavelet power between responders and nonresponders (relative to nonresponders), $p$-values correspond to the Mann-Whitney U test.

|  | EO | | | | | EC | | | | | | | |
| --- | --- | --- | --- | --- | --- | --- | --- | --- | --- | --- | --- | --- | --- |
| Channel | Fz | F4 | C3 | C4 | P3 | F4 | Fpz | Fz | F3 | C3 | P3 | P4 | Oz |
| AUC | 0.74 | 0.83 | 0.84 | 0.74 | 0.84 | 0.76 | 0.87 | 0.79 | 0.74 | 0.73 | 0.76 | 0.76 | 0.79 |
| $p$-value | 0.11 | 0.03 | 0.02 | 0.11 | 0.02 | 0.09 | 0.001 | 0.06 | 0.11 | 0.13 | 0.09 | 0.09 | 0.06 |
| Cut-off point | 0.029 | 0.03 | 0.024 | 0.024 | 0.041 | 0.029 | 0.041 | 0.029 | 0.031 | 0.017 | 0.039 | 0.0319 | 0.095 |
| Δ [%] | -26 | -20 | -29 | -18 | -26 | -11 | -29 | -20 | -13 | -29 | -31 | -28 | 39 |
| Sensitivity [%] | 70 | 90 | 80 | 70 | 70 | 70 | 80 | 70 | 90 | 60 | 90 | 60 | 90 |
| Specificity [%] | 71 | 71 | 86 | 71 | 100 | 85 | 86 | 90.9 | 57 | 86 | 71 | 86 | 71 |
| Accuracy [%] | 71 | 82 | 82 | 71 | 82 | 77 | 82 | 71 | 77 | 71 | 82 | 71 | 82 |

Table 3 The outcome of binary classification (prediction of treatment response) based on the normalized alpha wavelet power for eyes-open (EO) and eyes-closed (EC) condition. The power was calculated for the LM montage. Only the channels for which the AUROC was significantly greater than 0.5 are presented. Δ is the relative difference of the wavelet power between responders and nonresponders (relative to nonresponders), $p$-values correspond to the Mann-Whitney U test.

|  | EO | | | | | EC | | | | | | | | |
| --- | --- | --- | --- | --- | --- | --- | --- | --- | --- | --- | --- | --- | --- | --- |
| Channel | Fz | T3 | C3 | T4 | T6 | Fp1 | Fpz | F3 | Fz | T3 | C3 | O1 | O2 | Oz |
| AUC | 0.79 | 0.76 | 0.76 | 0.77 | 0.74 | 0.76 | 0.74 | 0.83 | 0.80 | 0.83 | 0.83 | 0.86 | 0.80 | 0.81 |
| $p$-value | 0.06 | 0.09 | 0.07 | 0.07 | 0.10 | 0.09 | 0.11 | 0.03 | 0.04 | 0.03 | 0.03 | 0.01 | 0.04 | 0.03 |
| Cut-off point | 0.056 | 0.029 | 0.047 | 0.027 | 0.024 | 0.0491 | 0.054 | 0.056 | 0.056 | 0.03 | 0.054 | 0.054 | 0.05 | .049 |
| Δ [%] | -9 | -23 | -18 | -22 | -23 | -16 | -14 | -16 | -15 | -24 | -18 | 77 | 61 | 84 |
| Sensitivity [%] | 70 | 90 | 70 | 90 | 80 | 70 | 60 | 80 | 60 | 90 | 90 | 80 | 70 | 80 |
| Specificity [%] | 100 | 71 | 86 | 71 | 71 | 71 | 71 | 71 | 86 | 71 | 71 | 86 | 86 | 71 |
| Accuracy [%] | 82 | 82 | 76 | 82 | 76 | 71 | 65 | 76 | 71 | 82 | 82 | 82 | 77 | 76 |

Table 4 The outcome of binary classification (prediction of treatment response) based on the normalized alpha wavelet power for eyes-open condition. The power was calculated for REF montage. Only the channels for which the AUROC was significantly greater than 0.5 are presented. Δ is the relative difference of the wavelet power between responders and nonresponders (relative to nonresponders), *p*-values correspond to the Mann-Whitney U test.

| Channel | F4 | C3 | C4 |
|---|---|---|---|
| AUC | 0.76 | 0.73 | 0.73 |
| *p*-value | 0.08 | 0.13 | 0.13 |
| Cut-off point | 0.0432 | 0.0165 | 0.0247 |
| Δ [%] | -25 | -36 | -23 |
| Sensitivity [%] | 100 | 80 | 70 |
| Specificity [%] | 71 | 71 | 86 |
| Accuracy | 88 | 76 | 76 |

Table 5 The response index (the percentage of channels in which a patient is classified as responder) for the ARE and LM montages. The index was calculated for eyes-open (EO) and eyes-closed (EC) conditions. The average value of the index for both conditions (EO+EC) is also presented.

| | RESPONDERS | | | | NONRESPONDERS | | |
|---|---|---|---|---|---|---|---|
| ID | RESPONSE [%] | | | ID | RESPONSE [%] | | |
| | EO | EC | EO+EC | | EO | EC | EO+EC |
| | | | ARE | | | | |
| R1 | 100 | 87.5 | 93.75 | N1 | 0 | 0 | 0 |
| R2 | 100 | 87.5 | 93.75 | N2 | 0 | 12.5 | 6.25 |
| R3 | 60 | 75 | 67.5 | N3 | 0 | 25 | 12.5 |
| R4 | 80 | 100 | 90 | N4 | 40 | 12.5 | 26.25 |
| R5 | 100 | 100 | 100 | N5 | 20 | 12.5 | 16.25 |
| R6 | 20 | 50 | 35 | N6 | 40 | 50 | 45 |
| R7 | 60 | 50 | 55 | N7 | 40 | 50 | 45 |
| R8 | 60 | 87.5 | 73.75 | | | | |
| R9 | 80 | 87.5 | 83.75 | | | | |
| R10 | 100 | 37.5 | 68.75 | | | | |
| | | | LM | | | | |
| R1 | 100 | 89 | 94 | N1 | 20 | 22 | 21 |
| R2 | 100 | 100 | 100 | N2 | 0 | 0 | 0 |
| R3 | 40 | 33 | 37 | N3 | 60 | 11 | 35 |
| R4 | 60 | 100 | 80 | N4 | 60 | 0 | 30 |
| R5 | 100 | 100 | 100 | N5 | 0 | 33 | 17 |
| R6 | 80 | 22 | 51 | N6 | 0 | 11 | 6 |
| R7 | 100 | 67 | 83 | N7 | 0 | 89 | 44 |
| R8 | 60 | 100 | 80 | | | | |
| R9 | 60 | 100 | 80 | | | | |
| R10 | 100 | 44 | 72 | | | | |

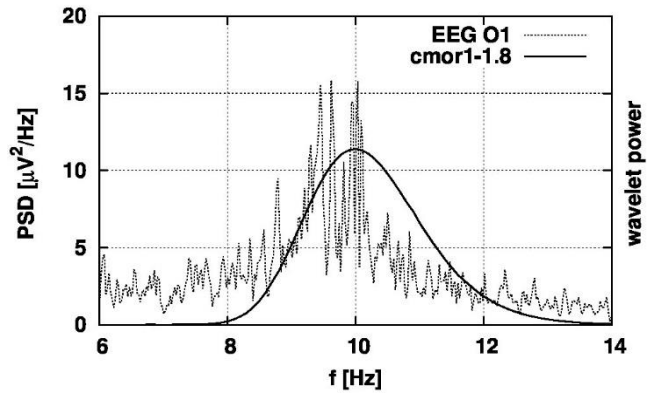

Fig. 1 The wavelet power of the monochromatic signal with frequency 10 Hz (solid line). The power was calculated using Morlet mother function ($f_c$=1.8 and $f_b$=1). The example of power spectral density of patient's EEG (channel O1) is shown with the dotted line.

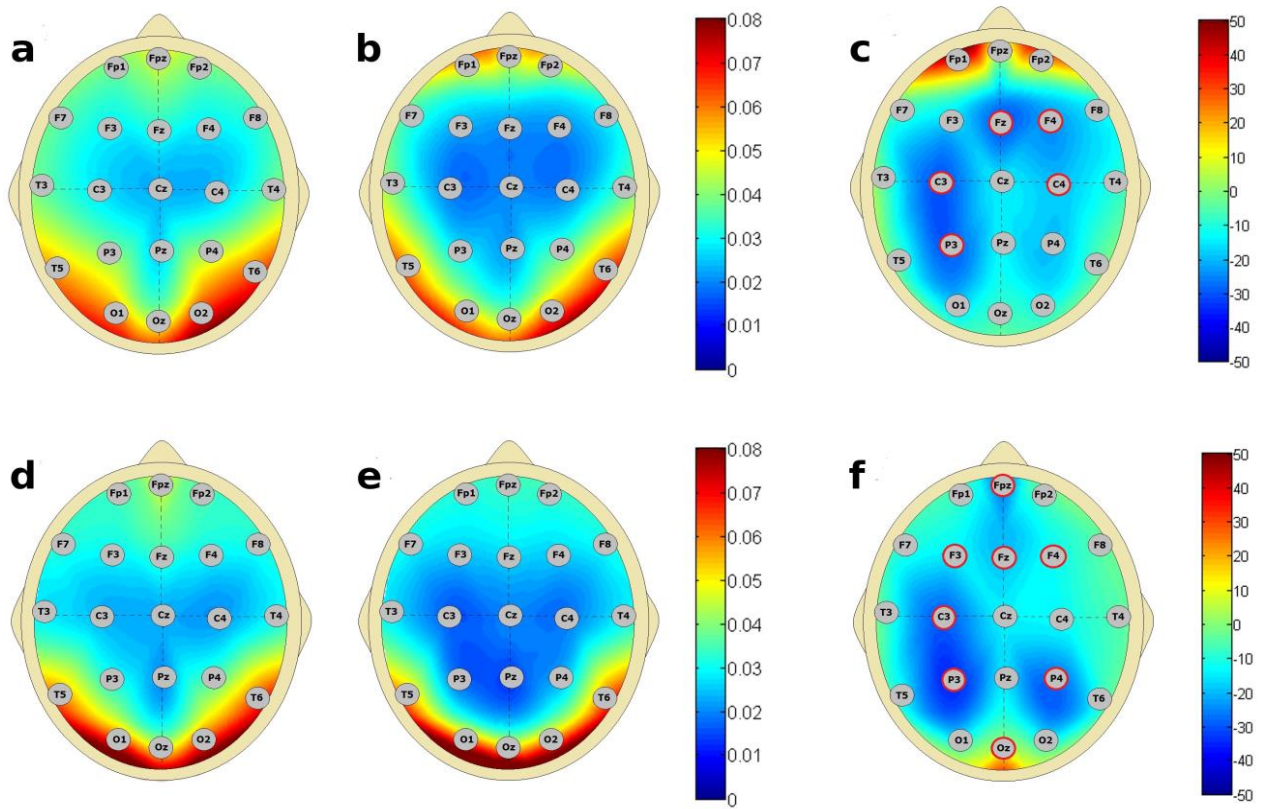

Fig. 2 Topography of alpha wavelet power for nonresponders (a and d) and responders (b and e) in eyes open (first row) and eyes closed conditions (second row). Figs. c and f show the relative difference of the wavelet power between responders and nonresponders (relative to nonresponders) for open and closed eyes, respectively. The red thick circles around EEG site labels indicate channels for which the AUROC was significantly greater than 0.5.

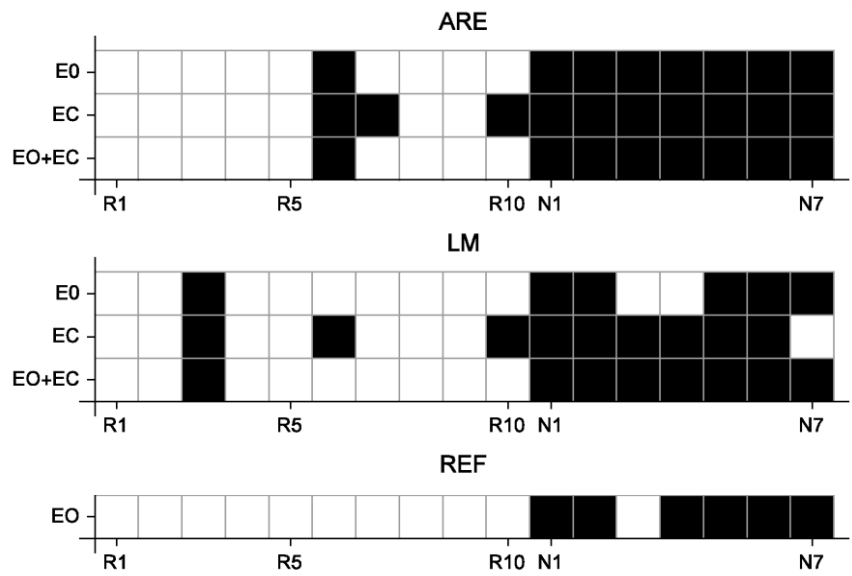

Fig. 3 Outcome of the prediction of antidepressant treatment response for three different montages: ARE (top), LM (middle), and REF (bottom). For ARE and LM montages, the prediction was based on the value of the response index for open (EO) and closed (EC) eyes, and the average over both conditions. For REF montage the prediction was possible only for open eyes. White and black boxes represent assignment to responders (R) and nonresponders (N), respectively.